\documentclass{iau}
\usepackage{graphicx}

\def\be{\begin{equation}}
\def\ee{\end{equation}}
\def\ba{\begin{eqnarray}}
\def\ea{\end{eqnarray}}
\def\go{\mathrel{\raise.3ex\hbox{$>$}\mkern-14mu
             \lower0.6ex\hbox{$\sim$}}}
\def\lo{\mathrel{\raise.3ex\hbox{$<$}\mkern-14mu
             \lower0.6ex\hbox{$\sim$}}}

\def\cE{{\cal E}}
\def\cR{{\cal R}}
\def\cI{{\cal I}}
\def\bmu{{\mbox{\boldmath $\mu$}}}
\def\bphi{{\mbox{\boldmath $\phi$}}}

\title[IAUS291.~~Merging neutron star binaries] 
{Merging neutron star binaries: equation of state and electrodynamics}

\author[D. Lai]  
{Dong Lai}
\affiliation{Department of Astronomy, Cornell University, Ithaca, NY 14853, USA
\\ email: {\tt dong@astro.cornell.edu}}

\pubyear{2012}
\volume{291}  
\jname{\mbox{Neutron Stars and Pulsars: Challenges and Opportunities after 80 years}}
\editors{J. van Leeuwen, ed.} 
\begin{document}

\maketitle

\begin{abstract}
Merging neutron star (NS) binaries may be detected by ground-based
gravitational wave (GW) interferometers (e.g. LIGO/VIRGO) within this
decade and may also generate electromagnetic radiation detectable
by wide-field, fast imaging telescopes that are coming online. 
The GWs can provide 
new constraint on the NS equation of state (including mass-radius relation 
and the related nuclear symmetry energy through resonant g-modes). 
This paper reviews various
hydrodynamical (including equilibrium and dynamical/resonant tides)
and electrodynamical processes in coalescing NS
binaries, with focus on the pre-merger phase.  
\end{abstract}


\firstsection 
\section{Introduction}

I was asked to talk about thermal radiation from
isolated neutron stars (NSs). In this meeting, George Pavlov
reviewed the X-ray properties of pulsars and thermally emitting NSs
(see Kaplan et al.~2011), and Wynn Ho discussed central compact
objects and their magnetic fields (see Halpern \& Gotthelf 2010;
Shabaltas \& Lai 2012; Vigano \& Pons 2012). Recent works on
theoretical modelling of NS surface emission can be found in Potekhin
et al.~(2012) (see also Pavlov et al.~1995; Harding \& Lai 2006 and
van Adelsberg \& Lai 2006 for reviews). Since these subjects were
adequately covered in the meeting, I decided to focus on a different
topic that did not receive much attention in this meeting but is
likely to become increasingly important in the coming decade.

Merging NS binaries have been studied since 1970s, with major
activities in the relativity community since the early 1990s because
of their importance as a source of gravitational waves (GWs)
(e.g. Cutler et al.~1993).  They are of great current interest for two
reasons: (i) Merging NS/NS or NS/Black-Hole (BH) binaries have been
identified as the leading candidate for the central engine of short
GRBs
(Berger 2011).  They are also expected to produce optical and radio
transients that may be detected by wide-field, fast imaging telescopes
that are coming online (e.g. PTF, LSST) in the next few years
(Nissanke et al.~2012).  (ii) After several decades of developmemt and
promise,
gravitational wave astronomy in the Hz-kHz band may finally take off
in the next decade. The initial LIGO reached the design sensitivity
($h_c\simeq 10^{-21}$) in 2006, and the enhanced LIGO (with a factor of 2
reduction in $h_c$) is taking or analysing data. The Advanced LIGO and
VIRGO are expected to begin observations in 2015 and reach full
sensitivity (a factor of 10 reduction in $h_c$) in 2018-19 --- at which
time the detection of GWs from many merging NS binaries seems
guaranteed.

The last three minutes of a NS binary's life may be divided into two
phases: the inspiral phase, producing quasi-periodic GWs, and the
coalescence phase, where physical collision results in ``messy'' GWs.
The recent years, 3D simulations of the final merger in full general
relativity (GR) have become possible (see Shibata \& Taniguchi 2006;
Foucart et at.~2012; Sekiguchi et al.~2012). It has long been
recognized that the final merger waveforms can provide a useful probe of NS
equation of state (EOS; e.g., Cutler et al.~1993; Bildsten \& Cutler
1992; Lai \& Wiseman 1996; Wiggins \& Lai 2000). The idea is 
simple: By measuring the ``cut-off'' frequency $\propto
(GM_t/R^3)^{1/2}$ associated with binary contact or tidal disruption,
combined with the precise mass measurement from the inspiral waveform,
one can obtain the NS radius (see Bauswein et al.~2012 for recent 
simulations which put such a idea into concrete footing; see
also Sekiguchi et al.~2012; Faber \& Rasio 2012 for reviews). 

In the following sections I will focus on the pre-merger phase.

\section{Hydrodynamics of merging NS binaries}

Prior to binary merger, tidal effects may affect the orbital decay and
the GWs. There are two types of tides: {\it equilibrium tides} and
{\it dynamical tides}. The equilibrium tides correspond to global deformation 
of the NS, which leads to the interaction potential between the two stars
(with the NS mass $M$ and radius $R$, the companion mass $M'$ -- treated as a 
point mass, and the binary separation $a$) 
\be
V(r)=-{MM'/a}-{\cal O}\left(k_2{{M'}^2R^5/a^6}\right),
\ee
where $k_2$ is the so-called
Love number. This would lead to a correction to the number of GW cycles,
$dN=dN^{(0)}\left[1-{\cal O}(k_2M'R^5/Ma^5)\right]$.  For a Newtonian
polytropic NS model, simple analytic expressions can be found in Lai et
al.~(1994). Recent semi-analytic GR calculations of such equilibrium
tidal effects (including the more precise determination of the Love number)
can be found in numerous papers (e.g., Flanagan \& Hinderer 2008;
Binnington \& Poisson 2009; Damour \& Nagar 2009; Penner et al.~2012, 
Ferrari et al.~2012). Obviously
this effect is only important at small orbital separations (just prior
to merger) -- there is some prospect of measuring this, thereby
constraining the EOS, but it may be challenging (Damour et al.~2012).
At small orbital separations, the quadrupole approximation is not valid; 
therefore one
must use the numerically computed GR quasi-equilibrium binary sequences
to characterize the tidal effect -- such sequences have been
constructed by several groups since the 1990s (e.g., Baumgarte et al.~1998;
Uryu et al.~2009) or use fully numerical simulations.

Another aspect of the equilibrium tide concerns tidal dissipation,
which leads to a lag of the tidal bulge with respect to the binary
axis.  It was shown already in the 1990s (Bildsten \& Cutler 1992;
Kochanek 1992) that because of the rapid GW-driven orbital decay,
viscous tidal lag cannot synchronize the NS spin. Thus the NS will
be close to irrotational (approximated as a Riemann-S ellipsoid;
Lai et al.~1994; Wiggins \& Lai 2000; Ferrari et al.~2012). 
Near the final phase
of the inspiral, the rapid orbital decay gives rise to a finite lag
angle (even with zero viscosity), but this cannot synchronize the NS
(Lai \& Shapiro 1995; Dall'Osso \& Rossi 2012).

The situation is more complicated for {\bf dynamical tides}, which
manifest as resonant excitations of internal oscillations of the NS: As
two NSs spiral in, the orbit can momentarily come into resonance with
the normal modes (frequency $\omega_\alpha$) of the NS:
\be
\omega_\alpha=m\Omega_{\rm orb},\qquad m=2,3,\cdots
\ee
By drawing energy from the orbital
motion and resonantly exciting the modes, the rate of inspiral is
modified, giving rise to a phase shift in the gravitational
waveform. This problem was studied by Reisenegger \& Goldreich (1994),
Lai (1994) and Shibata (1994) in the case of non-rotating NSs, where
the only modes that can be resonantly excited are g-modes (with
typical mode frequencies$\lo 100$~Hz). It was found that the effect is
small for typical NS parameters (mass $M=1.4M_\odot$ and radius
$R=10$~km) because the coupling between the g-mode and the tidal
potential is weak. Ho \& Lai (1999) studied the effect of NS rotation,
and found that the g-mode resonance can be strongly enhanced even by a
modest rotation (e.g., the phase shift in the waveform $\Delta\Phi$
reaches up to 0.1~radian for a spin frequency $\nu_s\lo 100$~Hz).
They also found that for a rapidly rotating NS ($\nu_s\go 500$~Hz),
f-mode resonance becomes possible (since the inertial-frame f-mode
frequency can be significantly reduced by rotation) and produces a
large phase shift. In addition, NS rotation gives rise to r-mode
resonance whose effect is appreciable only for very rapid (near
breakup) rotations. Lai \& Wu (2006) further studied resonant
excitations of other inertial modes (of which r-mode is a member) and
found similar effects.  Flanagan \& Racine (2006) studied the
gravitomagnetic resonant excitation of r-modes and and found that the
post-Newtonian effect is more important than the Newtonian tidal
effect (and that the phase shift reaches 0.1~radian for $\nu_s\sim
100$~Hz). Tsang et al.~(2012) examined crustal modes and found that 
the GW phase correction is small/modest and suggested that tidal resonance
could shatter the NS crust, giving rise to the pre-cursor of short GRBs.
Taken together, these studies suggest that for canonical NS
parameters ($R \simeq 10$~km, $\nu_s\lo 100$~Hz), tidal resonances
have a small effect on the gravitational waveform during binary
inspiral.  However, it is important to keep in mind that the effect is 
a strong function of $R$ (e.g., $\Delta\Phi\propto R^4$ for g-modes
and $\propto R^{3.5}$ for inertial modes). A larger radius ($R\simeq 
15$~km), appropriate for stiff EOS, 
would make the effect important. In the case of g-modes, the magnitude
of the effect depends on the symmetry energy of nuclear matter 
and could be non-negligible (W. Newton \& D. Lai 2013, in prep).

\section{Electrodynamics of merging NS binaries}

For magnetic NSs, magnetic interactions may play a role.  If the
binary is embedded in a vacuum, then the interaction potential is
$V(r)=-MM'/a-{\cal O}(\mu\mu'/a^3)$ (where $\mu,\mu'$ are the magnetic
dipole moments of the two stars). It is easy to check that such
magnetic interaction would lead to negligible effect on the GWs unless
both NSs have superstrong fields ($\gg 10^{15}$~G) -- this is unlikely
(e.g., the double pulsars PSR J0737-3039 has $10^{10}$~G for pulsar A
and $2\times 10^{13}$ for pulsar B). 

\begin{figure}[b]
\vspace*{-0.8 cm}                                                             
\begin{center}
 \includegraphics[width=3.4in]{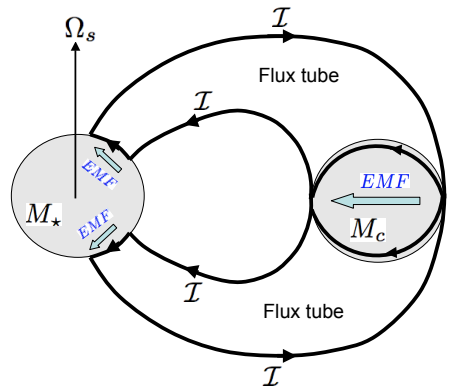}
\vspace*{-1.0 cm}                                                             
 \caption{DC circuit (unipolar induction) 
model of magnetic interactions in binary 
systems {\it a la} Goldreich \& Lynden-Bell (1969).}
   \label{fig1}
\end{center}
\end{figure}

Of course, as in the case of isolated pulsars,
the circumbinary environment cannot be vacuum.
The following discussion is based on Lai (2012).
Consider a binary system consisting of a magnetic NS (the
``primary'', with mass $M$, radius $R$, spin $\Omega_s$, and magnetic
dipole moment $\mu$) and a non-magnetic companion (mass $M_c$, radius
$R_c$). The orbital angular
frequency is $\Omega$.  The magnetic field strength at the surface of
the primary is $B_\star=\mu/R^3$.  The whole binary system is embedded
in a tenuous plasma (magnetosphere).  We assume for simplicity that
${\bf\Omega}$, ${\bf\Omega_s}$ and $\bmu$ are all aligned.
The motion of the non-magnetic companion relative to the magnetic field of the
primary produces an EMF $\cE \simeq 2R_c |E|$, where ${\bf E}=                         
{\bf v}_{\rm rel}\times {\bf B}/c$, with
${\bf v}_{\rm rel}=(\Omega-\Omega_s)a\,{\hat\bphi}$ and
${\bf B}=(-\mu/a^3){\hat{\bf z}}$.
This gives
$\cE\simeq {(2\mu R_c/ca^2)}\Delta\Omega$,
where $\Delta\Omega=\Omega-\Omega_s$\footnote{If the magnetic field does
have time to fully penetrate the companion star due to the rapid orbital decay,
then the EMF will be reduced by a factor of order the ratio of the skin depth
and the stellar radius. I thank Anatoly Spitkovsky for
pointing this out to me at the IAU conference.}. 
The EMF drives a current along the magnetic
field lines in the magnetosphere, connecting the primary and the companion
through two flux tubes. The current in the circuit is given by
$\cI={\cE/(\cR_{\rm tot})}$,
where the total resistance of the circuit is
$\cR_{\rm tot}=\cR+\cR_c+2\cR_{\rm mag}$,
with $\cR,\,\cR_c,\,\cR_{\rm mag}$ the resistances of the magnetic
star, the companion and the magnetosphere, respectively.
These resistances depend on the properties of the binary components and the
magnetosphere, and can vary widely for different types of systems.
The energy dissipation rate of the system is then
$\dot E_{\rm diss}=2{\cal I}^2R_{\rm tot}={2\cE^2/\cR_{\rm tot}}$,
where the factor of 2 accounts for both the upper and lower sides of the circuit.

The total magnetic force (in the azimuthal direction) on the companion is
$F_\phi\simeq (2R_c) (2{\cal I}B_z/c)$, with $B_z=-\mu/a^3$.
Thus the torque acting on the binary's orbital angular momentum is
$T=\dot J_{\rm orb}\simeq ({4/c})a\,R_c{\cal I}B_z\simeq
-({4\mu R_c/ca^2})({\cE/{\cal R}_{\rm tot}})$.
The torque on the primary's spin is $I\dot\Omega_s=-T$ (where $I$ is
the moment of inertia).
The orbital energy loss rate associated with $T$ is then
$\dot E_{\rm orb}=T\Omega$.

The equations above show that the binary
interaction torque and energy dissipation associated with the DC
circuit increase with decreasing total resistance $\cR_{\rm tot}$.
Is there a problem for the DC model when $\cR_{\rm tot}$ is too small?
The answer is yes.
The current in the circuit produces a toroidal magnetic field, which
has the same magnitude but opposite direction above and below
the equatorial plane. The toroidal field just above the
companion star (in the upper flux tube) is $B_{\phi+}\simeq -(2\pi/c){\cal J}_r$,
where ${\cal J}_r\simeq -4\cI/(\pi R_c)$ is the (height-integrated) surface current.
Thus the azimuthal twist of the flux tube is
$\zeta_\phi=-{B_{\phi+}/B_z}=
={16 v_{\rm rel}/(c^2\cR_{\rm tot})}$,
where $v_{\rm rel}=a\Delta\Omega=a(\Omega-\Omega_s)$.
Clearly, when $\cR_{\rm tot}$ is less than $16v_{\rm rel}/c^2$,
the flux tube will be highly twisted.

Goldreich \& Lynden-Bell (1969) speculated that the DC circuit would break
down when the twist is too large. (For the Jupiter-Io system
parameters,  the twist $|\zeta_\phi|\ll 1$.) 
Numerous works have since confirmed that this is indeed the case.
Theoretical studies and numerical simulations, usually carried out
in the contexts of solar flares
and accretion disks, have shown that as a flux tube is twisted beyond
$\zeta_\phi\go 1$, the magnetic pressure associated with $B_\phi$
makes the flux tube expand outward and the magnetic fields open up,
allowing the system to reach a lower energy state (e.g., Aly 1985; 
Lynden-Bell \& Boily 1994;
Lovelace et al.~1995; Uzdensky et al.~2002).
Thus, a DC circuit with $\zeta_\phi\go 1$ cannot be realized: The
flux tube will break up, disconnecting the linkage between the two
binary components.
A binary system with $\cR_{\rm tot}\lo 16v_{\rm rel}/c^2$
cannot establish a steady-state DC circuit.
The electrodynamics is likely rather complex, only
a quasi-cyclic circuit may be possible (Lai 2012; see Aly \& Kuijpers 1990):
(a) The magnetic field from the primary penetrates
part of the companion, establishing magnetic linkage between the two
stars; (b) The linked fields are twisted by differential rotation, generating
toroidal field from the linked poloidal field; (c) As the toroidal magnetic field
becomes comparable to the poloidal field, the fields inflate and
the flux tube breaks, disrupting the magnetic linkage;
(d) Reconnection between the inflated field lines relaxes the shear and restore
the linkage. The whole cycle repeats.

In general, we can use the dimensionless azimuthal twist $\zeta_\phi$
to parameterize the magnetic torque and energy dissipation rate:
\be
T= {1\over 2}aR_c^2B_zB_{\phi+}
= -\zeta_\phi{\mu^2 R_c^2\over 2a^5},\quad
\dot E_{\rm diss} = -T \Delta\Omega
=\zeta_\phi\Delta\Omega {\mu^2 R_c^2\over 2a^5}.
\label{eq:emax}\ee
The maximum torque and dissipation are obtained by setting $\zeta_\phi\sim 1$.
If the quasi-cyclic circuit discussed in the last paragraph is
established, we would expect $\zeta_\phi$ to vary between $0$ and $\sim 1$.
Note that in the above, $T$ is negative since we
are assuming $\Omega>\Omega_s$. 

Gravitational wave (GW) emission drives the orbital decay of the 
NS binary, with timescale
$t_{\rm GW}={a/|\dot a|}=
0.012\left({a/30\,{\rm km}}\right)^{4}{\rm s}$,
where we have adopted $M=1.4M_\odot$ and mass ratio
$q=M_c/M=1$.
The magnetic torque tends to spin up the primary when $\Omega$$>$$\Omega_s$.
Spin-orbit synchronization is possible only if
the synchronization time $t_{\rm syn}=I\Omega/|T|$ is less than
$t_{\rm GW}$ at some orbital radii.  With
$I=\kappa M R^2$, we find
\be
t_{\rm syn}={2\kappa(1+q)\over\zeta_\phi\Omega}
\left(\!{GM^2\over B_\star^2R^4}\!\right)
\!\left(\!{a\over R_c}\!\right)^2
\simeq 2\times 10^7\zeta_\phi^{-1}\!\left(\!{B_\star\over 10^{13}\,{\rm G}}
\!\right)^{\!-2}\!\left({a\over 30\,{\rm km}}\right)^{7/2}{\rm s},
\label{eq:tsyn}\ee
where on the right we have adopted $\kappa=0.4$ and $R=R_c=10$~km.
Clearly, even with magnetar-like field strength ($B_\star\sim 10^{15}$~G) and
maximum efficiency ($\zeta_\phi\sim 1$), spin-orbit synchronization cannot be
achieved by magnetic torque. For the same reason, the effect of magnetic torque on the
number of GW cycles during binary inspiral is small. 

The energy dissipation rate is
\be
\dot E_{\rm diss}=\zeta_\phi\left(\!{v_{\rm rel}\over c}\!\right){B_\star^2R^6
R_c^2c\over 2a^6}
= 7.4\times 10^{44}\zeta_\phi\left(\!{B_\star\over
10^{13}\,{\rm G}}\!\right)^{\!2}\!\left(\!{a\over 30\,{\rm km}}\!\right)^{\!\!-13/2}
\!{\rm erg\,s}^{-1},
\ee
where on the right we have used $v_{\rm rel}\simeq a\Omega$ (for
$\Omega_s\ll \Omega$) and adopted canonical parameters
($M=M_c=1.4M_\odot$, $R=R_c=10$~km).
The total energy dissipation per $\ln a$ is
\be
{dE_{\rm diss}\over d\ln a}=\dot E_{\rm diss}t_{\rm GW}
\simeq 8.9\times 10^{42}\zeta_\phi\!\left(\!{B_\star\over
10^{13}\,{\rm G}}\!\right)^{\!2}\!\left({a\over 30\,{\rm km}}\right)^{\!\!-5/2}
\!{\rm erg}.
\ee
Some fraction of this dissipation will emerge as electromagnetic
radiation counterpart of binary inspiral. It is possible that 
this radiation is detectable at extragalactic distance. But this will depend 
on the microphysics in the magnetosphere, including particle acceleration and 
radiation mechanism (e.g., Vietri 1996; Hansen \& Lyutikov 2001).

If one assumes that the magnetosphere resistance is given by the
impedance of free space, $\cR_{\rm mag}=4\pi/c$, then the corresponding twist
is $\zeta_\phi=2v_{\rm rel}/(\pi c)$, which satisfies our upper limit.
We then have
\be
\dot E_{\rm diss}=\left(\!{v_{\rm rel}\over c}\!\right)^2\!
{B_\star^2R^6 R_c^2c\over \pi a^6}
= 1.7\times 10^{44}\left(\!{B_\star\over 10^{13}\,{\rm G}}\!\right)^{\!2}
\!\left({a\over 30\,{\rm km}}\right)^{\!\!-7}{\rm erg/s}.
\ee
This is in agreement with the estimate of Lyutikov (2011).

The situation is similar for NS/BH binaries.
In the membrane paradigm (Thorne et al.~1986), a BH
of mass $M_H$ resembles a sphere of radius $R_c=R_H=2GM_H/c^2$
(neglecting BH spin)
and impedance $\cR_H=4\pi/c$. Neglecting the resistances of the
magnetosphere and the NS, the azimuthal twist of the flux tube in the DC
circuit is
$\zeta_\phi={4v_{\rm rel}/(\pi c)}$,
which satisfies our upper limit.
The energy dissipation rate is (cf. Lyutikov 2011; McWilliams \& Levin 2011)
\be
\dot E_{\rm diss}=\left(\!{v_{\rm rel}\over c}\!\right)^2\!
{2B_\star^2R^6 R_H^2c\over \pi a^6}
\simeq 5.7\!\times\! 10^{42}\!\left(\!{B_\star\over 10^{13}\,{\rm G}}\!\right)^{\!2}
\!\!\left(\!{M_H\over 10M_\odot}\!\right)^{\!\!\!-4}
\!\!\!\left(\!{a\over 3R_H}\!\right)^{\!\!-7}\!\!\!{\rm erg\,s}^{-1},
\ee
where we have assumed $M_{BH}/M\gg 1$.
Again, it is uncertain whether this radiation is detectable for binaries
at extragalactic distances.

\vspace{1ex}
{\bf Acknowledgements}:
This work has been supported in part by NSF grants AST-1008245 and
AST-1211061, and NASA grants NNX12AF85G and NNX10AP19G.

\end{document}